\documentclass[letterpaper,english,pra,preprint,superscriptaddress,nofootinbib]{revtex4}
\pdfoutput=1
\usepackage[T1]{fontenc}
\usepackage[latin9]{inputenc}
\usepackage{float}
\usepackage{amsmath}
\usepackage{graphicx}
\usepackage{amssymb}
\usepackage{esint}

\makeatletter
\@ifundefined{textcolor}{}
{%
 \definecolor{BLACK}{gray}{0}
 \definecolor{WHITE}{gray}{1}
 \definecolor{RED}{rgb}{1,0,0}
 \definecolor{GREEN}{rgb}{0,1,0}
 \definecolor{BLUE}{rgb}{0,0,1}
 \definecolor{CYAN}{cmyk}{1,0,0,0}
 \definecolor{MAGENTA}{cmyk}{0,1,0,0}
 \definecolor{YELLOW}{cmyk}{0,0,1,0}
 }

\makeatother

\makeatother

\usepackage{babel}

\makeatother

\usepackage{babel}

\begin{document}

\preprint{CERN-PH-TH/2009-248}

\preprint{MIT-CTP 4100}

\title{A Quantum Monte Carlo Method at Fixed Energy}

\author{Edward Farhi}

\email{farhi@mit.edu}

\affiliation{Center for Theoretical Physics, \\
Massachusetts Institute of Technology, Cambridge, MA 02139}

\author{Jeffrey Goldstone}

\email{goldston@mit.edu}

\affiliation{Center for Theoretical Physics, \\
Massachusetts Institute of Technology, Cambridge, MA 02139}

\author{David Gosset}

\email{dgosset@mit.edu}

\affiliation{Center for Theoretical Physics, \\
Massachusetts Institute of Technology, Cambridge, MA 02139}

\author{Harvey B. Meyer}

\email{meyerh@mit.edu}

\affiliation{Center for Theoretical Physics, \\
Massachusetts Institute of Technology, Cambridge, MA 02139}

\affiliation{Physics Department, CERN, 1211 Geneva 23, Switzerland}
\begin{abstract}
In this paper we explore new ways to study the zero temperature limit
of quantum statistical mechanics using Quantum Monte Carlo simulations.
We develop a Quantum Monte Carlo method in which one fixes the ground
state energy as a parameter. The Hamiltonians we consider are of the
form $H=H_{0}+\lambda V$ with ground state energy $E$. For fixed
$H_{0}$ and $V$, one can view $E$ as a function of $\lambda$ whereas
we view $\lambda$ as a function of $E$. We fix $E$ and define a
path integral Quantum Monte Carlo method in which a path makes no
reference to the times (discrete or continuous) at which transitions
occur between states. For fixed $E$ we can determine $\lambda(E)$
and other ground state properties of $H$. 
\end{abstract}
\maketitle

\section{Introduction}

Quantum Monte Carlo methods are widely used to compute properties
of quantum systems using classical sampling algorithms. In this paper
we develop a novel Quantum Monte Carlo method that allows one to numerically
investigate ground state properties of a quantum system.

A virtue of Quantum Monte Carlo is that one is not required to manipulate
vectors in the Hilbert space corresponding to the quantum system.
The dimension of this Hilbert space typically grows exponentially
with the physical size of the system. Instead, Quantum Monte Carlo
methods map the problem of approximating the ground state energy (or
some other observable) onto the problem of evaluating an expectation
value with respect to a probability distribution $q(X)$ over a set
of configurations $C$ (so $X\in C$). In order to evaluate this expectation
value, one can use a classical Markov chain Monte Carlo algorithm
to sample configurations from the distribution $q$. Markov chain
Monte Carlo works by defining a Markov chain on the space of configurations
$C$. This Markov chain can be described by an update rule which tells
you how to generate a new configuration of the chain from the current
one. The Markov chain is constructed so that the limiting distribution
is $q(X)$. One then applies some large number $N_{0}$ of iterations
of the Markov chain to some initial configuration $X_{0}$. If $N_{0}$
is sufficiently large then after these iterations, the distribution
of subsequent configurations will be arbitrarily close to $q$.

We now give a brief description of how our method is used to estimate
properties of the ground state. We write the Hamiltonian as $H(\lambda)=H_{0}+\lambda V$,
where $H_{0}$ is diagonal in a given basis $\{|z\rangle\}$ and $\lambda V$
is off diagonal in this basis. (In an n spin system $z$ is an n bit
string.) In section \ref{sec:The-Hamiltonian} we outline our assumptions
and restrictions on $H_{0}$ and $V$. With these choices, the ground
state energy is always less than or equal to zero, and we will see
that for each value of $E<0,$ there exists one positive value $\lambda(E)$
such that the ground state of $H(\lambda(E))$ has energy E. To use
our Monte Carlo method, one first must fix $E<0$ and a large integer
$m$. We define a path of length $m$ to be a sequence $\{z_{1},...,z_{m}\}$,
where each $n$ bit string $z_{i}$ is the label of the state $|z_{i}\rangle$.
These paths are the configurations of the previous paragraph. We will
define a probability distribution $f(\{z_{1},...,z_{m}\})$ over the
set of all paths of length m (this distribution is also a function
of the value of $E$ which was chosen). We will show how the function
$\lambda(E)$ can be obtained by computing an average with respect
to the probability distribution $f$.

To motivate our method and to get a general idea of how it works,
consider the function

\begin{equation}
G(E,\lambda)=Tr\bigg[\bigg(\frac{-\lambda}{H(\lambda)-E}\bigg)V\bigg]\,.\label{eq:G}\end{equation}
 Assuming that $E<0$ and $\lambda>0$ are chosen so that the Taylor
series expansion converges, we can write\begin{eqnarray}
G(E,\lambda) & = & Tr\bigg[\bigg(\frac{-\lambda}{1+\frac{\lambda}{H_{0}-E}V}\bigg)\frac{1}{H_{0}-E}V\bigg]\nonumber \\
 & = & \sum_{m=1}^{\infty}Tr\bigg[\bigg(\frac{-\lambda}{H_{0}-E}V\bigg)^{m}\bigg]\,.\label{eq:taylor}\end{eqnarray}
 It is clear from the expression in equation \ref{eq:G} that the
function $G(E,\lambda)$ blows up when $E\rightarrow E_{g}(\lambda)$,
where $E_{g}(\lambda)$ is the ground state energy of $H(\lambda)$.
Equivalently we can say that at a fixed value of $E$ the blow up
occurs as $\lambda\rightarrow\lambda(E)$, where $E_{g}(\lambda(E))=E.$
At this value of $\lambda$ the Taylor series expansion must diverge.
In fact, this divergence occurs because as $m$ becomes large, terms
in the series approach 1 for large $m$ (here we have made some assumptions
about the Hamiltonian which we discuss in the next section) so \[
Tr\bigg[\bigg(\frac{-\lambda(E)}{H_{0}-E}V\bigg)^{m}\bigg]\approx1\,.\]
 For the remainder of this section we assume that $m$ is large enough
to make $\approx$ close to $=$. By inserting complete sets of states
in the basis that diagonalizes $H_{0}$ we can express the LHS as
a sum over paths \[
(\lambda(E))^{m}\sum_{\{z_{1},...,z_{m}\}}\langle z_{1}|-V|z_{m}\rangle\langle z_{m}|-V|z_{m-1}\rangle...\langle z_{2}|-V|z_{1}\rangle\prod_{i=1}^{m}\frac{1}{E_{i}-E}\approx1\]
 where $E_{i}=\langle z_{i}|H_{0}|z_{i}\rangle$. Now taking the log
and differentiating with respect to $E$, we obtain\begin{eqnarray}
-\frac{1}{\lambda(E)}\frac{d\lambda(E)}{dE} & \approx & \sum_{\{z_{1},...,z_{m}\}}f\left(\{z_{1},...,z_{m}\}\right)\bigg(\frac{1}{m}\sum_{i=1}^{m}\frac{1}{E_{i}-E}\bigg)\nonumber \\
 & = & \langle\frac{1}{m}\sum_{i=1}^{m}\frac{1}{E_{i}-E}\rangle_{f}\,.\label{eq:firstderivmotivation}\end{eqnarray}
 Here the expectation value is taken with respect to the measure $f$
on paths defined by\[
f\left(\{z_{1},...,z_{m}\}\right)=\frac{1}{F}\langle z_{1}|-V|z_{m}\rangle\langle z_{m}|-V|z_{m-1}\rangle...\langle z_{2}|-V|z_{1}\rangle\prod_{i=1}^{m}\frac{1}{E_{i}-E}\]
 where $F$ is a normalizing constant. We will show how to sample
with respect to the distribution $f$ in a way that makes numerical
work possible. Sampling from the distribution $f$ will also allow
us to compute $-\frac{1}{\lambda(E)}\frac{d\lambda(E)}{dE}$ from
equation \ref{eq:firstderivmotivation} as well as other properties
of the ground state.

Our paper is organized as follows. In section \ref{sec:The-Hamiltonian}
we describe the types of Hamiltonians for which our method applies.
In section \ref{sec:A-Different-Ensemble} we outline the new method
that we propose. In section \ref{sec:An-Example:-Transverse} we explicitly
construct Monte Carlo update rules for the case where $V=-\sum_{i=1}^{n}\sigma_{x}^{i}$,
and we give numerical data using our algorithm at $n=16$ where we
are able to compare with exact diagonalization. In section \ref{sec:Standard}
we review the continuous imaginary time Quantum Monte Carlo method
\cite{prokofev-1996-64}, which is based on the thermal path integral.
We also derive a novel estimator in this ensemble of paths for the
ground state energy which becomes exact in the limit $\beta\rightarrow\infty.$

\section{The Hamiltonian\label{sec:The-Hamiltonian}}

We consider finite dimensional Hamiltonians of the form\[
H(\lambda)=H_{0}+\lambda V,\]
 where $H_{0}$ is diagonal in a given basis $\{|z\rangle\}$, and
$V$ has zeros along the diagonal in this basis. We make the following
assumptions about the Hamiltonian: 
\begin{enumerate}
\item The off diagonal matrix elements of $V$ in the basis $\{|z\rangle\}$
which diagonalizes $H_{0}$ are all either negative or zero. (This
ensures that our Quantum Monte Carlo method will not suffer from a
sign problem.) 
\item The ground state of $H(\lambda)$ is not degenerate for any value
of $\lambda\in(-\infty,\infty).$ 
\item The smallest eigenvalue of $H_{0}$ is zero. Note that this condition
can be fulfilled without loss of generality by adding a constant term
to the Hamiltonian. Writing $|z_{0}\rangle\in\{|z\rangle\}$ for the
unique state with $H_{0}|z_{0}\rangle=0$, we further require that
$V|z_{0}\rangle\ne0$ . (This implies that $|z_{0}\rangle$ is not
an eigenvector of $V$ since $\langle z_{0}|V|z_{0}\rangle=0$ follows
from assumption 1 above.) 
\end{enumerate}
We write $|\psi_{g}(\lambda)\rangle$ and $E_{g}(\lambda)$ for the
ground state eigenvector and ground state energy of $H(\lambda)$
. From second order perturbation theory in $\lambda$, we have that\begin{eqnarray}
\frac{d^{2}E_{g}}{d\lambda^{2}} & = & -2\sum_{z\neq z_{0}}\frac{|\langle z|V|z_{0}\rangle|^{2}}{\langle z|H_{0}|z\rangle}\nonumber \\
 & < & 0\label{eq:2deriv}\end{eqnarray}
 where the inequality is strict because $V|z_{0}\rangle\neq0$.

Using the fact that \begin{eqnarray}
\frac{dE_{g}}{d\lambda} & = & \langle\psi_{g}(\lambda)|V|\psi_{g}(\lambda)\rangle\label{eq:feynman-hellman}\end{eqnarray}
 we show that\[
\frac{dE_{g}}{d\lambda}=\begin{cases}
>0\text{\text{\text{ ,} for}\;\ensuremath{\lambda}<0}\\
=0\text{ , for}\;\lambda=0\\
<0\text{ , for\: \ensuremath{\lambda}>0}\,. & \text{}\end{cases}\]

\begin{figure}
\includegraphics[scale=0.5]{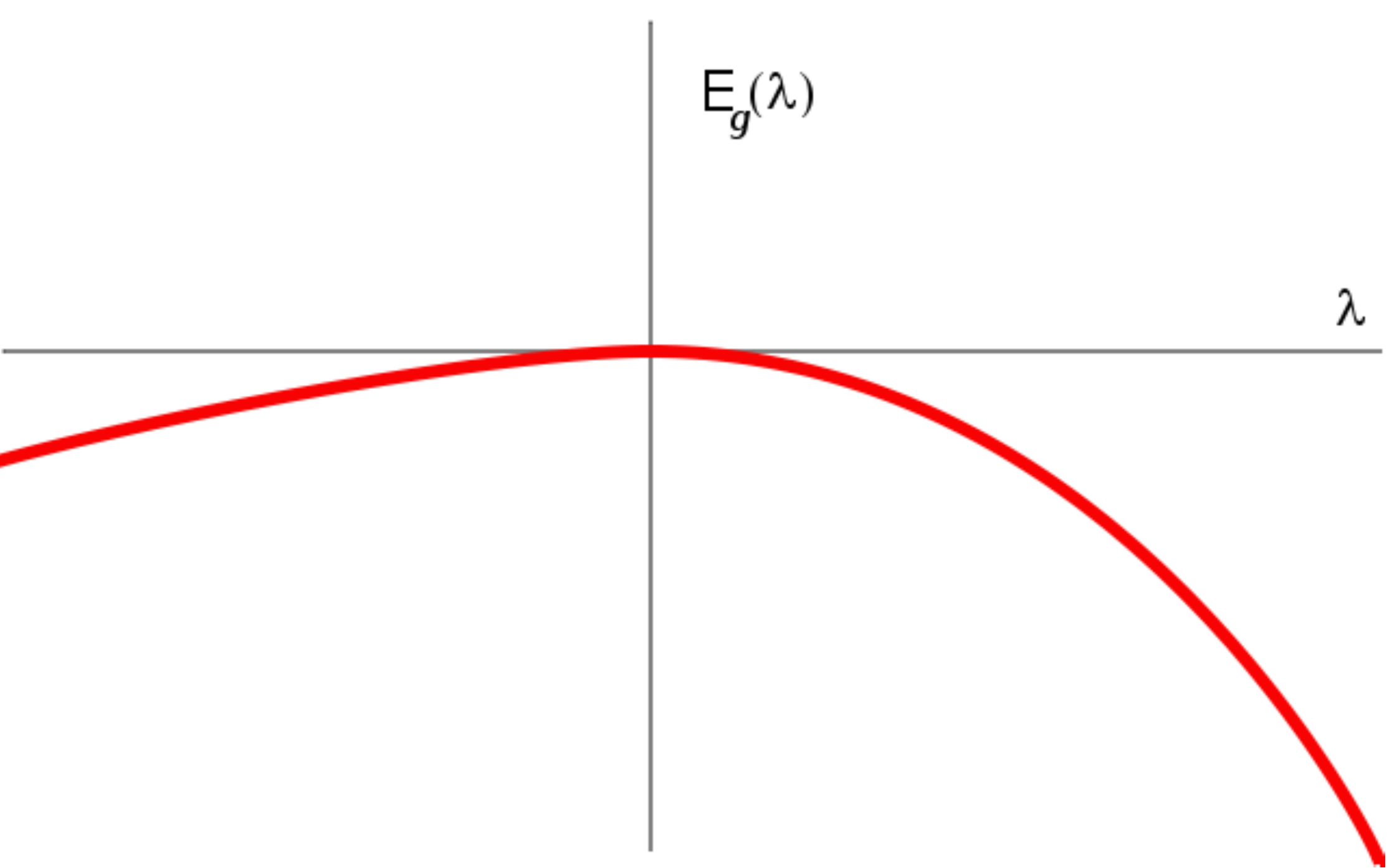}

\caption{$E_{g}(\lambda)$ for the Hamiltonians we consider. As $\lambda\rightarrow\pm\infty$
we have $E_{g}\rightarrow-\infty.$}

\label{Flo:eoflambda} 
\end{figure}

In order to obtain the inequalities, we use the variational principle.
When $\lambda>0,$ the ground state energy must be less than zero,
since $|z_{0}\rangle$ has zero expectation value for $H$ (and $|z_{0}\rangle$
is not an eigenvector of $H(\lambda)$). This, together with the fact
that $H_{0}$ is positive semidefinite, implies that $\langle\psi_{g}|V|\psi_{g}\rangle<0.$
The analogous result for $\lambda<0$ is obtained in the same way.
These inequalities give a qualitative picture of the curve $E_{g}(\lambda)$.
Starting from $E_{g}(0)=0,$ the curve slopes downwards as it goes
out from $\lambda$=0, and approaches $-\infty$ on both sides of
the origin for sufficiently large $|\lambda|.$ Note that this implies
that for each $E<0$ there is one positive and one negative value
of $\lambda$ (call them $\lambda(E)$ and $\lambda_{-}(E)$ respectively)
such that $E_{g}(\lambda(E))=E$ and $E_{g}(\lambda_{-}(E))=E$. Furthermore,
we show in appendix \ref{sub:A(E)_properties} that it is always the
case that\begin{equation}
\lambda(E)\leq|\lambda_{-}(E)|\,.\label{eq:lambdaineq}\end{equation}
We refer to the case where the inequality is strict as the generic
case. We illustrate the qualitative features of the curve $E_{g}(\lambda)$
(for the generic case) in figure \ref{Flo:eoflambda}. %
 In the nongeneric case where equality holds at some particular value
of $E$ , then in fact equality holds at every value of $E$ and the
curve $E_{g}(\lambda)$ is symmetric about $\lambda=0$.

\section{A New Quantum Monte Carlo Method\label{sec:A-Different-Ensemble}}

\subsection*{Definition of the Ensemble of Paths and Relevant Estimators}

As motivated in the Introduction, we now define an ensemble where
the configurations are sequences $\{z_{1},...,z_{m}\}$ (where each
$z_{i}$ is an $n$ bit string) , and we show how properties of the
ground state can be computed in this ensemble. We refer to the sequences
$\{z_{1},...,z_{m}\}$ as paths.

To begin, we fix $E<0$ and a large integer $m$ as parameters. As
in section \ref{sec:The-Hamiltonian}, we take $\lambda(E)$ to be
the positive value of $\lambda$ such that $H(\lambda)$ has ground
state energy $E$, with corresponding eigenvector $|\psi_{g}(\lambda(E))\rangle$.
We now describe how our method allows us to approximate $\lambda(E)$
and other properties of the ground state.

Recall from the Introduction that the probability distribution $f$
over paths is defined by \begin{equation}
f(z_{1},...,z_{m})\equiv\frac{1}{F(E,m)}\langle z_{1}|-V|z_{m}\rangle\dots\langle z_{2}|-V|z_{1}\rangle\prod_{i=1}^{m}\frac{1}{E_{i}-E}\label{eq:little_f}\end{equation}
 with%
\footnote{The nongeneric case where equality holds in equation \ref{eq:lambdaineq}
can arise when there exists a unitary transformation $U$ such that
$U^{\dagger}VU=-V$ and $U^{\dagger}H_{0}U=H_{0}$. In this case it
is seen from equation \ref{eq:trace_eqn} that $F(E,m)=0$ when $m$
is odd. In the nongeneric case $m$ must always be taken to be even.%
} \begin{align}
F(E,m) & \equiv\sum_{\{z_{1},...,z_{m}\}}\langle z_{1}|-V|z_{m}\rangle...\langle z_{2}|-V|z_{1}\rangle\prod_{i=1}^{m}\frac{1}{E_{i}-E}\label{eq:def_F}\\
 & =Tr\bigg[(\frac{-1}{H_{0}-E}V)^{m}\bigg]\,.\label{eq:trace_eqn}\end{align}
 As examples, we now define two quantities $\frac{\bar{\beta}(E,m)}{m}$
and $\bar{\lambda^{2}}(E,m)$ as ensemble averages with respect to
the distribution $f$ on paths \begin{eqnarray}
\bar{\beta}(E,m) & \equiv & \sum_{\{z_{1},...,z_{m}\}}f(\{z_{1},...,z_{m}\})\beta_{est}(\{z_{1},...,z_{m}\})=\langle\beta_{est}\rangle_{f}\nonumber \\
\overline{\lambda^{2}}(E,m) & \equiv & \sum_{\{z_{1},...,z_{m}\}}f(\{z_{1},...,z_{m}\})\lambda_{est}^{2}(\{z_{1},...,z_{m}\})=\langle\lambda_{est}^{2}\rangle_{f}\text{ }\label{eq:meanvals}\end{eqnarray}
 where we have defined the estimators (hence the subscript) \begin{eqnarray}
\beta_{est}(\{z_{1},...,z_{m}\}) & \equiv & \sum_{i=1}^{m}\frac{1}{E_{i}-E}\label{eq:beta_est}\\
\lambda_{est}^{2}(\{z_{1},...,z_{m}\}) & \equiv & \frac{1}{m}\sum_{i=1}^{m}\delta_{z_{i+2}z_{i}}(E_{i+1}-E)(E_{i}-E)\frac{1}{\langle z_{i}|V^{2}|z_{i}\rangle}\label{eq:meanvals2}\end{eqnarray}
 with $E_{i}=\langle z_{i}|H_{0}|z_{i}\rangle$, and $z_{m+1}=z_{1}$,
$z_{m+2}=z_{2}$. Our reason for using the symbol $\beta_{est}$ will
become clear in section \ref{sec:Standard} where we will discuss
its interpretation as an inverse temperature. We show in appendix
\ref{sec:Derivation-of-Estimators} that the ensemble averages $\frac{\bar{\beta}(E,m)}{m}$
and $\bar{\lambda^{2}}(E,m)$ correspond to properties of the quantum
ground state in the limit $m\rightarrow\infty$\begin{eqnarray}
\lim_{m\rightarrow\infty}\frac{\bar{\beta}(E,m)}{m} & = & -\frac{1}{\lambda(E)}\frac{d\lambda(E)}{dE}\,\label{eq:lim_beta}\\
 & = & -\frac{1}{\lambda(E)}\frac{1}{\langle\psi_{g}(E)|V|\psi_{g}(E)\rangle}\label{eq:betaoverm}\\
\lim_{m\rightarrow\infty}\overline{\lambda^{2}}(E,m) & = & (\lambda(E))^{2}\,.\label{eq:lim_lambda}\end{eqnarray}
 Equation \ref{eq:lim_beta} is equation \ref{eq:firstderivmotivation}
of the introduction and equation \ref{eq:betaoverm} follows from
\ref{eq:feynman-hellman}. One can also derive expressions for higher
derivatives of $\log(\lambda(E))$ as averages with respect to $f$.

\subsection*{Monte Carlo Simulation}

We propose to use the measure $f$ as the basis for Monte Carlo simulations.
In particular, our Monte Carlo algorithm begins by choosing $E<0$
and a large integer $m$ and then samples sequences of bit strings
from the distribution $f$. One can then use the estimators from equations
\ref{eq:meanvals} and \ref{eq:meanvals2} to evaluate the quantities
$\lambda(E)$ and $\langle\psi_{g}(\lambda(E))|V|\psi_{g}(\lambda(E))\rangle$,
using equations \ref{eq:betaoverm} and \ref{eq:lim_lambda} for the
limiting behaviour of these estimators. We do not construct a general
method for sampling from $f,$ instead we leave it to the reader to
construct such a method for the particular choice of $V$ at hand.
We do note that it is essential that whatever method is used conserves
the total number $m$ of transitions in the path. We now give an example
of this for a generic spin system with $V=-\sum_{i=1}^{n}\sigma_{x}^{i}$.

\section{An Example: Transverse Field Spin Hamiltonians\label{sec:An-Example:-Transverse} }

\begin{figure}
\begin{centering}
\includegraphics[scale=0.5]{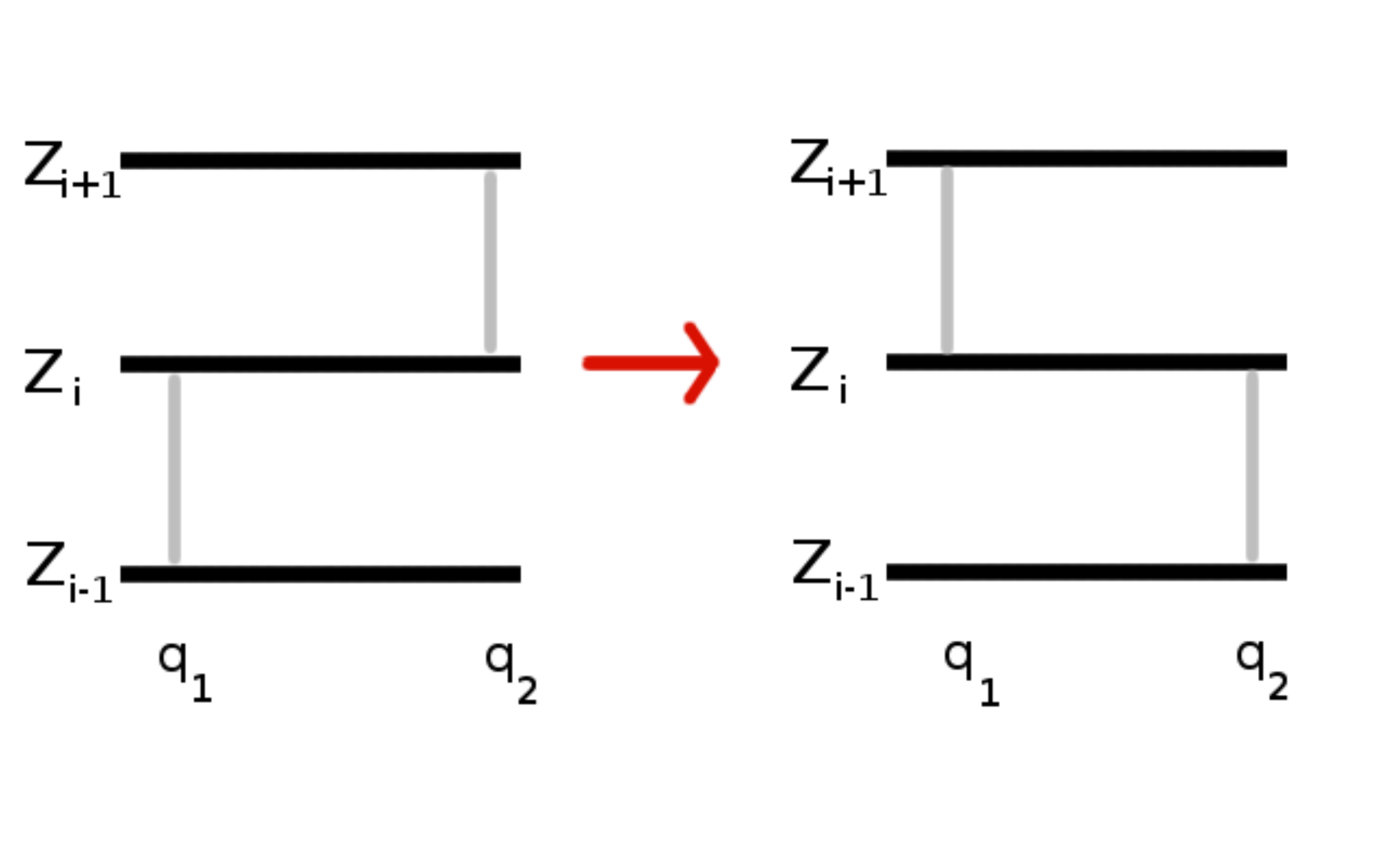} 
\par\end{centering}

\caption{Monte Carlo update where the order of 2 flips in the path is interchanged\label{fig:interchange_update}}

\end{figure}

We now explicitly construct a method to compute averages with respect
to the distribution $f$ in the case where the Hilbert space is that
of n spin $\frac{1}{2}$ particles, and $V=-\sum_{i=1}^{n}\sigma_{x}^{i}$.
The Hamiltonian $H_{0}$ is an arbitrary diagonal matrix in the Pauli
z basis for $n$ spins. Our algorithm can be used to compute the average
of any function of the path which is invariant under cyclic permutations
of the path. For this choice of $V$, the spectrum of $H(\lambda)$
is symmetric about $\lambda=0$ (so this corresponds to the nongeneric
case where equality holds in equation \ref{eq:lambdaineq} for all
$E<0$). With these choices, the paths which have nonzero weight (with
respect to $f$) are periodic paths of $m$ bit strings of length
$n$ where each string $z$ differs from the previous one by a bit
flip. We must take $m$ to be even since each bit must flip an even
number of times so that the path is periodic (and so the total number
of bit flips $m$ in the path must be even). In order to sample from
these paths according to $f$, we construct a Markov Chain which has
$f$ as its limiting distribution (actually, our Markov Chain converges
to the correct distribution over equivalence classes of paths which
are only defined up to cyclic permutation--this is why we restrict
ourselves to estimating quantities which are cyclically invariant).
Note that with our choice of $V$ we can write (see equation \ref{eq:little_f})
\[
f(\{z_{1},...,z_{m}\})=\frac{1}{F(E,m)}\prod_{i=1}^{m}\frac{1}{E_{i}-E}\,.\]
 Our Markov chain is defined by the following update rule which describes
how the configuration is changed at each step
\begin{enumerate}
\item Choose an integer $i\in\{1,...,m\}$ uniformly at random. 
\item Consider the bit strings $z_{i-1},z_{i},z_{i+1}$ in the current path
(where $z_{m+1}=z_{1}$). Suppose that $z_{i-1}$ and $z_{i}$ differ
in bit $q_{1}\in\{1,...,n\}$, which we write as $z_{i}=z_{i-1}\oplus\hat{e}_{q_{1}}$.
Also write $q_{2}\in\{1,...,n\}$ for the bit in which $z_{i}$ and
$z_{i+1}$ differ, so $z_{i+1}=z_{i}\oplus\hat{e}_{q_{2}}$. 
\item If $q_{1}\neq q_{2}$ then propose to change the bit string $z_{i}$
to the new value $\tilde{z}_{i}=z_{i-1}\oplus\hat{e}_{q_{2}}.$ Accept
this proposal with probability\[
P_{accept}=\min\left\{ 1,\frac{E_{i}-E}{\tilde{E}_{i}-E}\right\} \]
 where $\tilde{E}_{i}=\langle\tilde{z}_{i}|H_{0}|\tilde{z}_{i}\rangle$.
This Monte Carlo move has the effect of interchanging 2 consecutive
flips in the path (see figure \ref{fig:interchange_update}). 
\item If $q_{1}=q_{2}$ , then choose a new bit $q_{new}\in\{1,...,n\}$
from the probability distribution \begin{equation}
P(q_{new}=q)=\frac{1}{W}\frac{1}{E_{q}^{\prime}-E}\label{eq:same_q}\end{equation}
 where $E_{q}^{\prime}=\langle z_{i-1}\oplus\hat{e}_{q}|H_{0}|z_{i-1}\oplus\hat{e}_{q}\rangle$,
and $W=\sum_{j=1}^{n}\frac{1}{E_{j}^{\prime}-E}$. Then (with probability
1) change $z_{i}$ to the new value $z_{i-1}\oplus\hat{e}_{q_{new}}$.
This Monte Carlo move replaces a pair of consecutive flips which occur
in the same bit with 2 new flips in a possibly different bit (see
figure \ref{fig:replace_update}). 
\end{enumerate}
We show in appendix \ref{sec:Ergodicity-and-Detailed} that this algorithm
can be used to estimate any quantity which is invariant under cyclic
permutations of the path (note that all estimators we have discussed
have this property).

\begin{figure}
\begin{centering}
\includegraphics[scale=0.5]{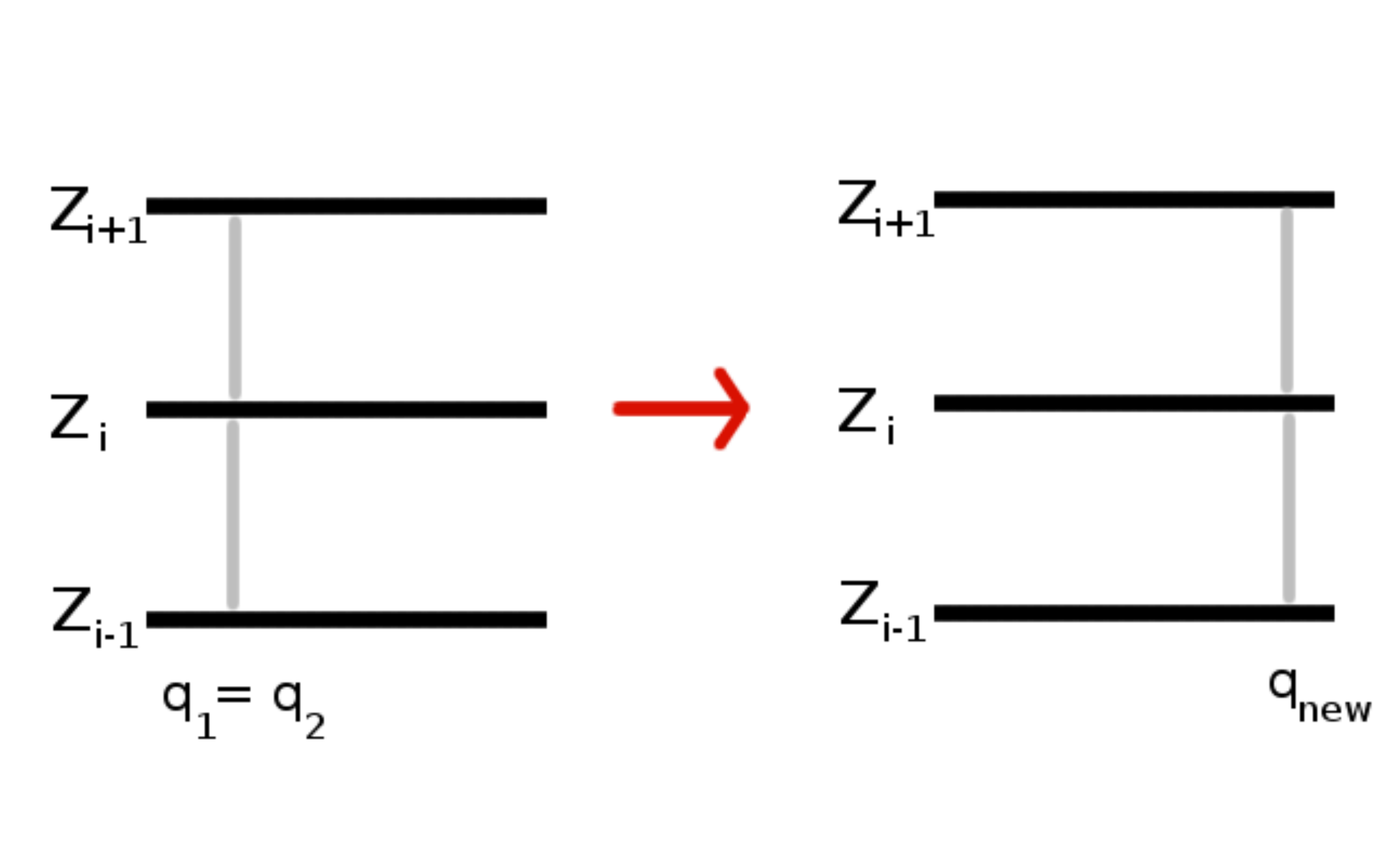} 
\par\end{centering}

\caption{Monte Carlo update where 2 adjacent flips in the path which occur
in the same bit are replaced by flips in a different bit.\label{fig:replace_update}}

\end{figure}

\subsection*{Numerical Simulation with a Particular Choice of $H_{0}$}

We have numerically tested our new Monte Carlo algorithm using a C++
computer program. In this section we show numerical data at 16 bits
where we are able to compare results with exact numerical diagonalization.
We studied the Hamiltonian with $V$ as in the previous section, and
$H_{0}$ corresponding to the combinatorial optimization problem Exact
Cover which stems from our interest in quantum computation \cite{farhi-2001}\[
H(\lambda)=H_{0}-\lambda\sum_{i=1}^{n}\sigma_{x}^{i}\]
 where\begin{eqnarray*}
H_{0} & = & \sum_{c=1}^{N_{c}}\bigg(\frac{1-\sigma_{z}^{i_{1}(c)}-\sigma_{z}^{i_{2}(c)}-\sigma_{z}^{i_{3}(c)}}{2}\bigg)^{2}\,.\end{eqnarray*}
 Here $H_{0}$ is a sum over $N_{c}$ terms, which in computer science
are called clauses. Each clause involves three distinct bits $i_{1}(c),i_{2}(c),i_{3}(c)$.
A clause is said to be satisfied by an n bit string $z$ if the state
$|z\rangle$ has zero energy for the corresponding term in the Hamiltonian.
The particular choice of $N_{c}$ and the bits involved in each clause
defines an instance of Exact Cover. Such an instance is said to be
satisfiable if there is an n bit string $z_{s}$ which satisfies all
clauses in the instance. In that case the z basis state $|z_{s}\rangle$
is the zero energy ground state of $H_{0}$, that is $H_{0}|z_{s}\rangle=0.$

\begin{figure}[H]

\begin{centering}
\includegraphics[scale=0.75]{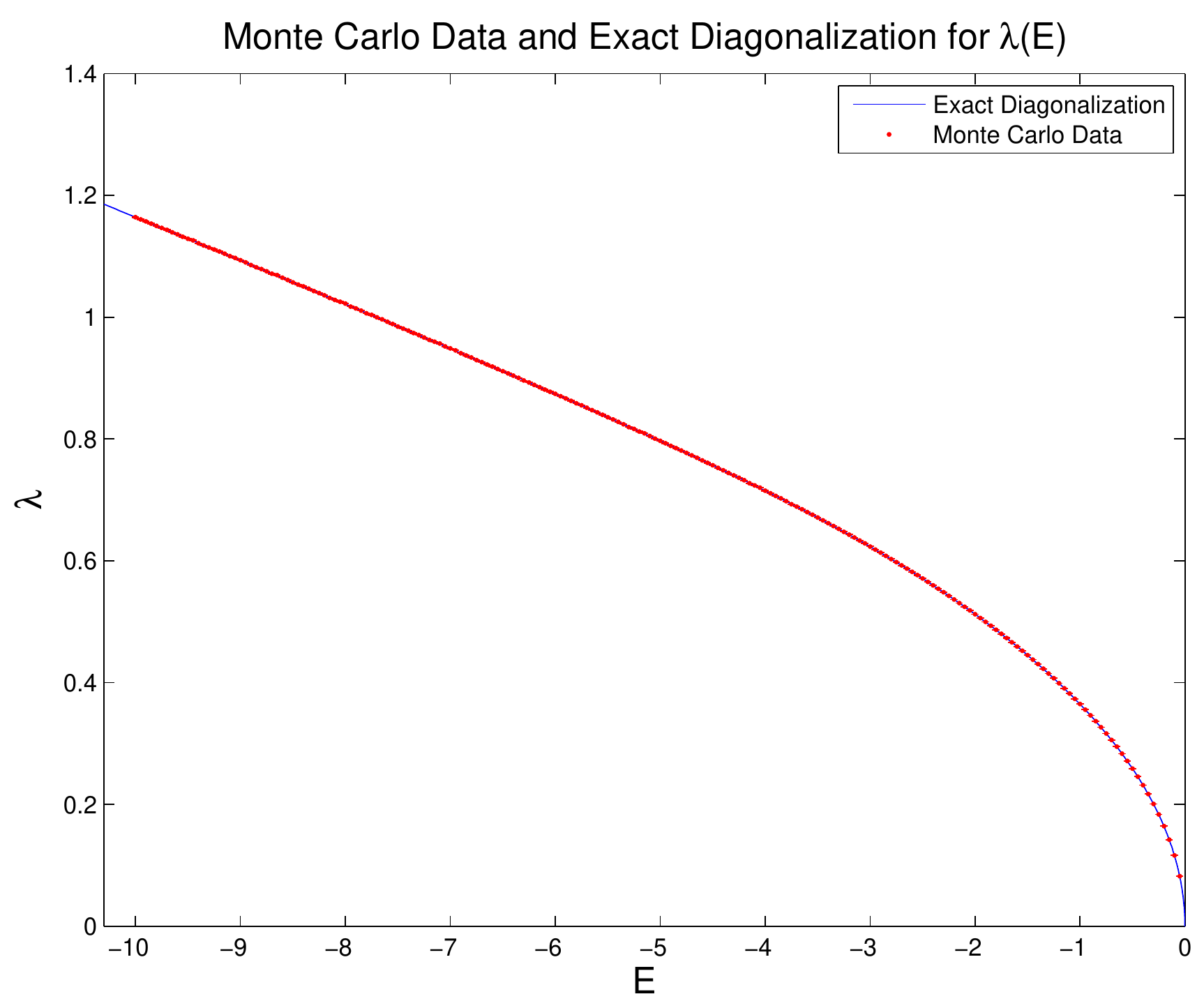} 
\par\end{centering}

\caption{$\lambda(E)$ computed using Monte Carlo data and exact diagonalization
for a 16 spin Hamiltonian. Statistical error bars are included for
the Monte Carlo results, but they are barely visible. We have also
plotted the errors separately in figure \ref{fig:errors_lambda}.}

\label{Flo:lambdaofE} 
\end{figure}

\begin{figure}[H]
 \protect

\begin{centering}
\protect\includegraphics[scale=0.75]{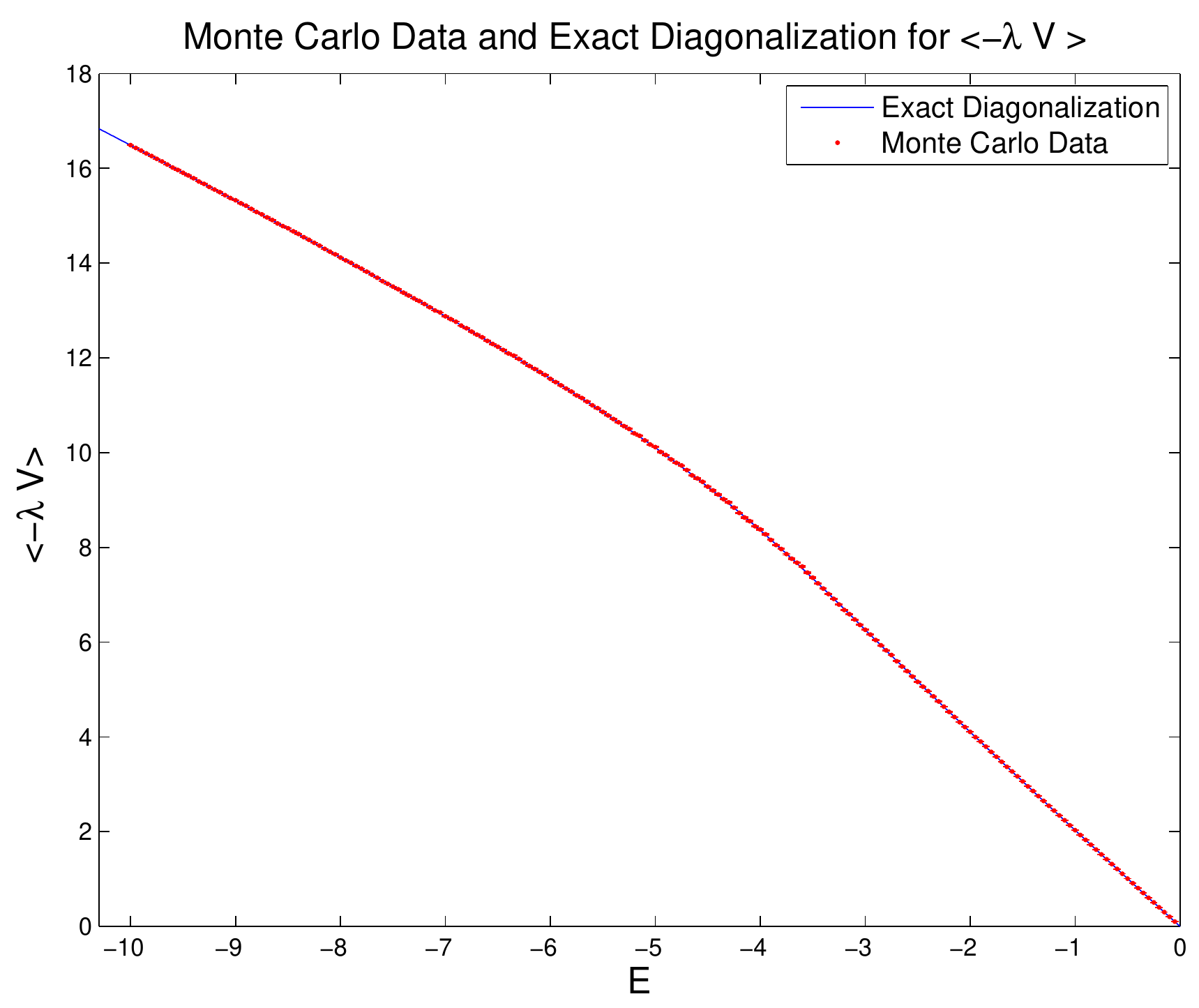} \protect 
\par\end{centering}

\caption{$-\lambda(E)\langle\psi_{g}(\lambda(E))|V|\psi_{g}(\lambda(E))\rangle$
computed using Monte Carlo data and exact diagonalization for a 16
spin Hamiltonian. Statistical error bars are included for the Monte
Carlo results, but they are barely visible. Errors are also plotted
separately in figure \ref{fig:errors_V}.}

\label{Flo:VofE} 
\end{figure}

\begin{figure}[H]
\protect

\begin{centering}
\protect\includegraphics[scale=0.5]{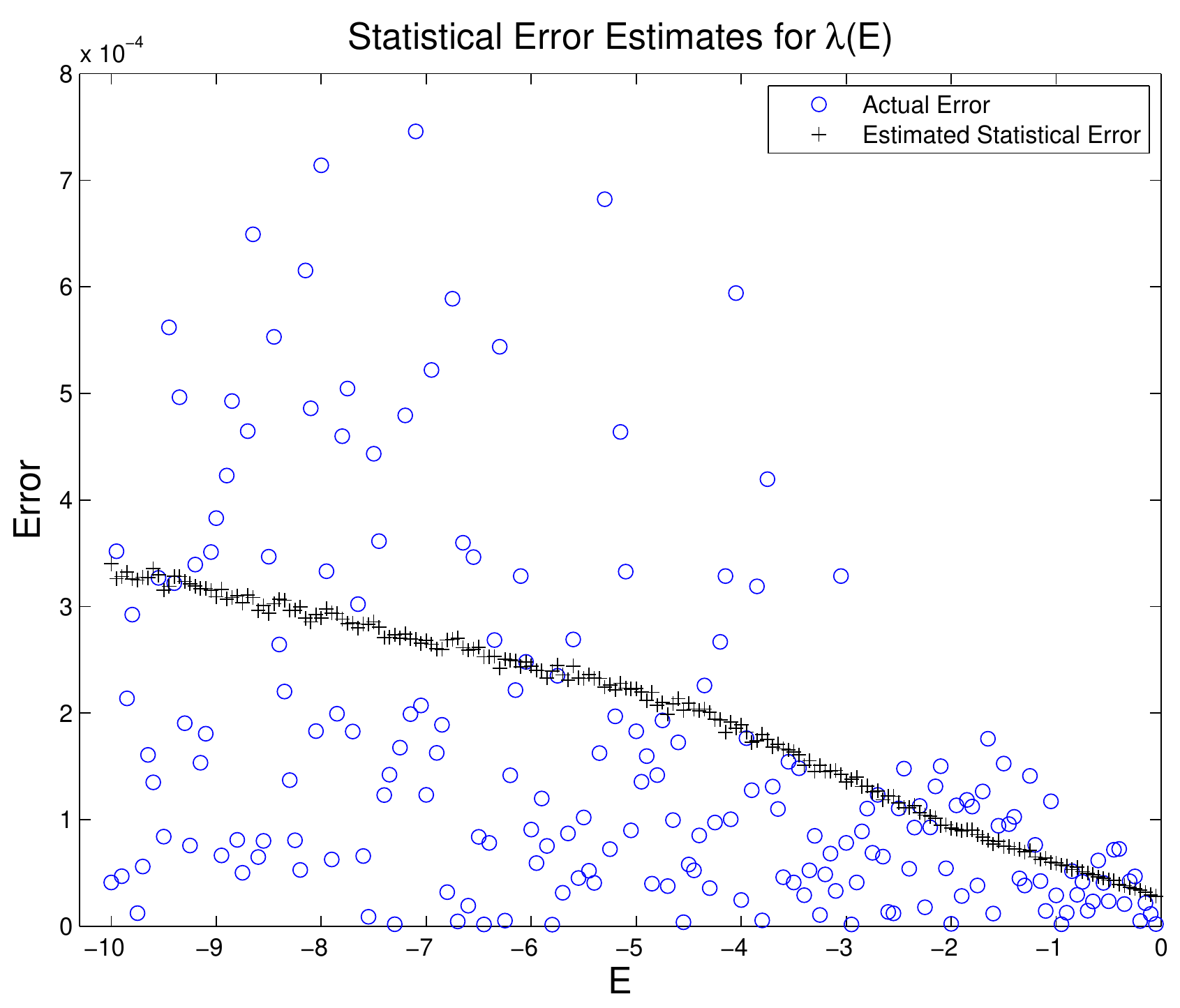}\protect 
\par\end{centering}

\caption{$\lambda(E)$ from figure \ref{Flo:lambdaofE}. The black crosses
show the estimated statistical error. The blue circles show the magnitude
of the difference between the Monte Carlo estimates and the result
of exact numerical diagonalization.\label{fig:errors_lambda}}

\end{figure}

\begin{figure}[H]
\protect

\begin{centering}
\protect\includegraphics[scale=0.5]{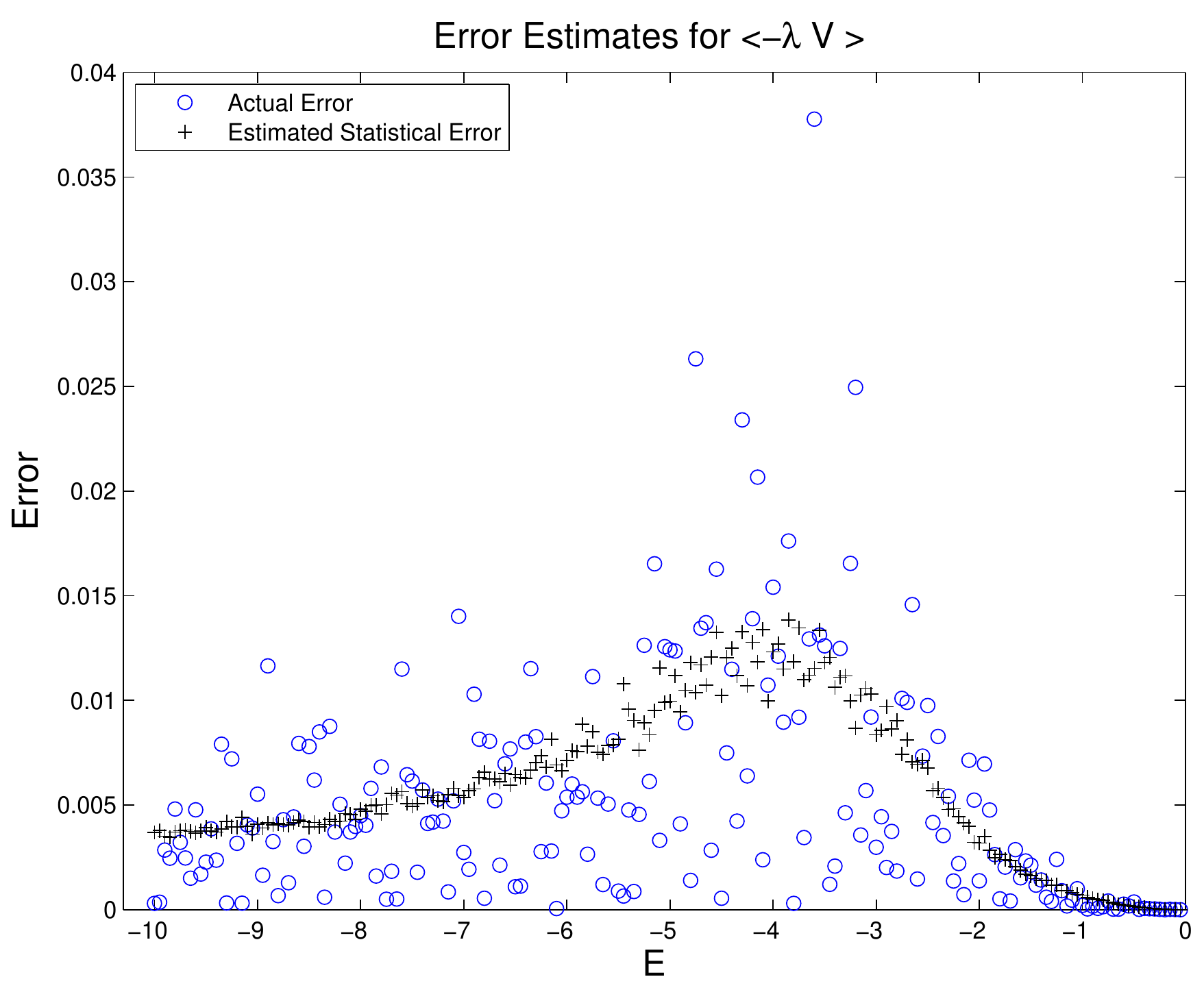}\protect 
\par\end{centering}

\caption{$-\lambda(E)\langle\psi_{g}(\lambda(E))|V|\psi_{g}(\lambda(E))\rangle$
from figure \ref{Flo:VofE}. The black crosses show the estimated
statistical error. The blue circles show the magnitude of the difference
between the Monte Carlo estimates and the result of exact numerical
diagonalization.\label{fig:errors_V}}

\end{figure}

We generated an instance of Exact Cover on 16 bits with a unique satisfying
assignment through a random procedure. Figures \ref{Flo:lambdaofE}
and \ref{Flo:VofE} show the values of $\lambda(E)$ and 
 $-\lambda(E)dE(\lambda)/d\lambda$ computed using equation \ref{eq:meanvals}
for $200$ values of $E$, with $m=1000$. Statistical errors were
computed using Ulli Wollf's error analysis program \cite{wolff-2004-383}.
This data set was taken by running $10^{8}$ Monte Carlo updates on
each of two processors of a dual core laptop computer for each value
of $E$. The two processors ran simultaneously and the total time
taken for all the data was under 5 hours. We also include the curves
for these quantities obtained by exact diagonalization, which are
in good agreement with the Monte Carlo data. Since it is hard to see
the error bars in figures \ref{Flo:lambdaofE} and \ref{Flo:VofE},
we have plotted the errors separately in figures \ref{fig:errors_lambda}
and \ref{fig:errors_V}.

\section{A New Estimator For The Energy in Standard Continuous Imaginary time
Quantum Monte Carlo \label{sec:Standard}}

In the ensemble of paths which was considered in the previous section,
the parameters $m$ and $E$ are fixed and then quantities $\bar{\beta}(E,m)$
and $\overline{\lambda^{2}}(E,m)$ are computed as ensemble averages.
In this section we review the standard approach to computing thermal
averages using continuous imaginary time Quantum Monte Carlo \cite{prokofev-1996-64}.
To use this method, parameters $\beta$ and $\lambda$ are fixed beforehand
and quantities $\bar{E}(\beta,\lambda)$ and $\bar{m}(\beta,\lambda)$
are computed as ensemble averages. We will also derive a new estimator
for $\langle H\rangle=\frac{Tr[He^{-\beta H}]}{Z(\beta)}$ which is
valid for large $\beta$ simulations. The form of this estimator establishes
a connection between the continuous imaginary time Quantum Monte Carlo
method and the new method that we outlined in the previous section.

The standard method \cite{prokofev-1996-64} is based on the expansion
of the partition function\begin{eqnarray}
Z(\beta) & = & Tr\bigg[e^{-\beta H}\bigg]\nonumber \\
 & = & Tr\bigg[\sum_{m=0}^{\infty}(-\lambda)^{m}e^{-\beta H_{0}}\int_{0}^{\beta}dt_{m}\int_{0}^{t_{m}}dt_{m-1}...\int_{0}^{t2}dt_{1}V_{I}(t_{m})V_{I}(t_{m-1})...V_{I}(t_{1})\bigg]\nonumber \\
 & = & Tr\bigg[e^{-\beta H_{0}}\bigg]+\sum_{m=1}^{\infty}\bigg[(-\lambda)^{m}\sum_{\{z_{1},...,z_{m}\}}\langle z_{1}|V|z_{m}\rangle\langle z_{m}|V|z_{m-1}\rangle...\langle z_{2}|V|z_{1}\rangle\nonumber \\
 &  & \int_{0}^{\beta}dt_{m}\int_{0}^{t_{m}}dt_{m-1}...\int_{0}^{t_{2}}dt_{1}e^{-(E_{1}t_{1}+E_{2}(t_{2}-t_{1})+...+E_{1}(\beta-t_{m}))}\bigg]\label{eq:part_func}\end{eqnarray}
 where $V_{I}(t)=e^{tH_{0}}Ve^{-tH_{0}}$ and $E_{i}=\langle z_{i}|H_{0}|z_{i}\rangle$.
This expression is interpreted as a path integral, where a path is
defined by a piecewise constant function $z(t)$ for $t\in[0,\beta]$.
The function $z(t)$ takes values in the set $\{z\}$ which are the
labels of the basis states $\{|z\rangle\}$ which diagonalize $H_{0}.$
In the above expression for $Z(\beta)$, we have\begin{eqnarray*}
z(t) & = & \begin{cases}
z_{1}, & 0\leq t<t_{1}\\
z_{2}, & t_{1}\leq t<t_{2}\\
\;\vdots\\
z_{m}, & t_{m-1}\leq t<t_{m}\\
z_{1}, & t_{m}\leq t\leq\beta\,.\end{cases}\end{eqnarray*}
 The allowed values of $m$ are $2,3,4,5,...$ as well as $m=0$ in
which case the path is constant $z(t)=z_{1}$ (for some $z_{1}$)
for all $t\in[0,\beta]$.

We define $H_{0}(z(t))$ to be $\langle z(t)|H_{0}|z(t)\rangle$,
so $\int_{0}^{\beta}H_{0}(z(t))dt=E_{1}t_{1}+E_{2}(t_{2}-t_{1})+...+E_{1}(\beta-t_{m})$
. Equation \ref{eq:part_func} can be used to define a measure $\rho$
on paths in imaginary time. The probability of a given path parameterized
by $z(t)$ is

\begin{eqnarray*}
\rho(z) & = & \begin{cases}
\frac{1}{Z(\beta)}e^{-\beta H_{0}(z_{1})}, & m=0\\
\frac{1}{Z(\beta)}\lambda^{m}\langle z_{1}|-V|z_{m}\rangle\langle z_{m}|-V|z_{m-1}\rangle...\langle z_{2}|-V|z_{1}\rangle e^{-\int_{0}^{\beta}H_{0}(z(t))dt}dt_{1}...dt_{m}, & m\neq0\end{cases}\end{eqnarray*}
 where $Z(\beta)$ is the normalization. Our assumption that off diagonal
elements of $V$ are nonpositive guarantees that $\rho\geq0$. The
task of sampling paths from the distribution $\rho$ can be accomplished
using Markov chain Monte Carlo methods such as those outlined in references
\cite{krzakala-2008-78,prokofev-1996-64,farhi-2009}. By using these
methods to sample paths from this probability distribution, one can
compute physical properties of the quantum system at an inverse temperature
$\beta.$ Known estimators for the expectation of the terms in the
Hamiltonian are (see appendix \ref{sec:Estimators-in-Continuous})

\begin{eqnarray}
\langle H_{0}\rangle & = & \frac{Tr[H_{0}e^{-\beta H}]}{Z(\beta)}=\langle\frac{1}{\beta}\int_{0}^{\beta}H_{0}(z(t))dt\rangle_{\rho}\label{eq:standardestimators}\\
\langle\lambda V\rangle & = & \frac{Tr[\lambda Ve^{-\beta H}]}{Z(\beta)}=-\langle\frac{m}{\beta}\rangle_{\rho}\,.\label{eq:standardV}\end{eqnarray}
 Here $m$ is the number of transitions in the path $z(t)$ and is
not fixed. Note that the estimator for $\langle\lambda V\rangle$
only involves the number of transitions in the path. The above two
estimators can be combined to obtain $\langle H\rangle$. One can
also obtain the following expressions for the variances of these estimators
in the limit $\beta\rightarrow\infty$

\begin{eqnarray*}
\beta{\rm Var}\bigg(\frac{1}{\beta}\int_{0}^{\beta}H_{0}(z(t))dt & \bigg)\stackrel{\beta\to\infty}{\to} & -\lambda^{2}\frac{d^{2}E_{g}}{d\lambda^{2}}\\
\beta{\rm Var}(\frac{m}{\beta}) & \stackrel{\beta\to\infty}{\to} & -\lambda\langle\psi_{g}|V|\psi_{g}\rangle-\lambda^{2}\frac{d^{2}E_{g}}{d\lambda^{2}}\,.\end{eqnarray*}
 This shows that as $\beta\rightarrow\infty$, the distributions for
these estimators become sharply peaked about their mean values. We
define intensive ensemble averages\begin{eqnarray*}
\frac{\bar{m}(\beta,\lambda)}{\beta} & \equiv & \langle\frac{m}{\beta}\rangle_{\rho}\\
\bar{E}(\beta,\lambda) & \equiv & \langle\frac{1}{\beta}\int_{0}^{\beta}H_{0}(z(t))dt\rangle_{\rho}-\frac{\bar{m}(\beta,\lambda)}{\beta}\end{eqnarray*}
 which in the $\beta\to\infty$ limit become respectively the ground
state expectation value of $-\lambda V$ and the ground state energy.

\subsection*{A New Estimator for the Ground State Energy}

We now derive the following novel estimator for the energy in the
standard ensemble, which is useful in the limit $\beta\rightarrow\infty$
:

\[
\langle H\rangle=\langle E^{\star}\rangle_{\rho}+O\left(\frac{1}{\beta}\right)\]
 where $E^{\star}$ is a function of the path defined to be the smallest
value of $E$ which satisfies the equation \begin{equation}
\beta=\sum_{i=1}^{m+1}\frac{1}{E_{i}-E}\,.\label{eq:eqstate}\end{equation}
 In this expression the $E_{i}$ are the energies of the states $|z_{i}\rangle$
visited along the path, with $E_{m+1}=E_{1}.$ Note that the above
equation is almost identical to equation \ref{eq:beta_est} (this
justifies our choice of notation $\beta_{est}$). We obtain this formula
by a similar method to that used in reference \cite{Beard20031012}
to obtain an alternate estimator for the energy $\langle H\rangle$.
Our formula, however, is valid when some or all of the $\{E_{i}\}$
are the same, and therefore resolves a serious difficulty encountered
in reference \cite{Beard20031012}. This may be of use in large $\beta$
Monte Carlo simulations, as an alternative to the standard estimator
for $\langle H\rangle$. We note in particular that this estimator
does not involve the times $\{t_{1},...,t_{m}\}$ in the path. 

Equation \ref{eq:eqstate} can be derived by considering the Laplace
transform of the partition function (with $s>-E_{g}$)

\begin{eqnarray*}
\int_{0}^{\infty}e^{-\beta s}Z(\beta)d\beta & = & Tr\bigg[\frac{1}{H+s}\bigg]\,.\end{eqnarray*}
 We can also express this as\begin{eqnarray*}
\int_{0}^{\infty}e^{-\beta s}Z(\beta)d\beta & = & \int_{0}^{\infty}d\beta e^{-\beta s}\sum_{m=0}^{\infty}\bigg[\lambda^{m}\sum_{\{z_{1},...,z_{m}\}}\langle z_{1}|-V|z_{m}\rangle\langle z_{m}|-V|z_{m-1}\rangle...\langle z_{2}|-V|z_{1}\rangle\\
 &  & \int_{0}^{\infty}du_{m+1}...\int_{0}^{\infty}du_{1}e^{-\sum_{i=1}^{m+1}E_{i}u_{i}}\delta(\beta-\sum_{i=1}^{m+1}u_{i})\bigg]\\
 & = & \sum_{m=0}^{\infty}\lambda^{m}\sum_{\{z_{1},...,z_{m}\}}\langle z_{1}|-V|z_{m}\rangle...\langle z_{2}|-V|z_{1}\rangle\prod_{i=1}^{m+1}\frac{1}{E_{i}+s}\,.\end{eqnarray*}
 Performing the inverse Laplace transform gives \[
Z(\beta)=\sum_{m=0}^{\infty}\lambda^{m}\sum_{\{z_{1},...,z_{m}\}}\langle z_{1}|-V|z_{m}\rangle\dots\langle z_{2}|-V|z_{1}\rangle\frac{1}{2\pi i}\int_{C}dse^{\beta s}\bigg(\prod_{i=1}^{m+1}\frac{1}{E_{i}+s}\bigg)\,.\]
 C is a contour in the complex plane which encircles the poles of
the integrand, which are located at $\{-E_{i}\}$. The expectation
value of the energy can then be expressed as \begin{eqnarray}
\frac{1}{Z(\beta)}Tr[He^{-\beta H}] & =- & \frac{1}{Z}\frac{dZ}{d\beta}\nonumber \\
 & = & \bigg<\frac{\frac{1}{2\pi i}\int_{C}(-s)e^{\beta s}\bigg(\prod_{i=1}^{m+1}\frac{1}{E_{i}+s}\bigg)ds}{\frac{1}{2\pi i}\int_{C}e^{\beta s}\bigg(\prod_{i=1}^{m+1}\frac{1}{E_{i}+s}\bigg)ds}\bigg>_{\rho}\,.\label{eq:s_star}\end{eqnarray}
 The complex function $h(s)=e^{\beta s}\bigg(\prod_{i=1}^{m+1}\frac{1}{E_{i}+s}\bigg)$
will in general have multiple saddle points along the real axis. We
can solve for the locations of these saddle points by writing \[
h(s)=e^{\beta g(s)}\]
where \[
g(s)=s-\frac{1}{\beta}\sum_{i=1}^{m+1}\log(E_{i}+s).\]
 Saddle points occur at real values $s^{\star}$ where $\frac{dg}{ds}(s^{\star})=0$
, which says that\[
\beta=\sum_{i=1}^{m+1}\frac{1}{E_{i}+s^{\star}}\,.\]

In the integrals in equation \ref{eq:s_star}, we can choose the contour
of integration along the curve of steepest descent through that saddle
point $s^{\star}$ which is largest (it is possible to show that one
can deform the contour to follow this curve without changing the encircled
poles).

Performing the integrals in equation \ref{eq:s_star} and letting
$E^{\star}=-s^{\star}$ we obtain

\[
\frac{1}{Z(\beta)}Tr[He^{-\beta H}]=\langle E^{\star}\rangle+O\left(\frac{1}{\beta}\right)\]
 where $E^{\star}$ is the smallest solution to equation \ref{eq:eqstate}.

\section{Conclusions}

We have outlined a new approach to Quantum Monte Carlo simulations
in which properties of the ground state of a quantum system are computed
at a fixed value of the ground state energy. We have confirmed the
validity of our method by performing a numerical simulation of a system
consisting of 16 spins. Our approach involves a path integral which
does not include any jump time variables, as in the stochastic series
expansion \cite{0305-4470-25-13-017,sandvik-1997-56}. We have also
obtained a new estimator for the ground state energy which is valid
in continuous imaginary time Quantum Monte Carlo simulations but which
does not involve the imaginary time variables. We hope that our method
will be applied to numerical simulations that go beyond the toy system
studied in this paper.

\section{Acknowledgements}

The authors thank Sam Gutmann for useful discussions. This work was
supported in part by funds provided by the U.S. Department of Energy
under cooperative research agreement DE-FG02-94ER40818, the W. M.
Keck Foundation Center for Extreme Quantum Information Theory, the
U.S Army Research Laboratory's Army Research Office through grant
number W911NF-09-1-0438, the National Science Foundation through grant
number CCF-0829421, and the Natural Sciences and Engineering Research
Council of Canada.

\bibliographystyle{plain} \bibliographystyle{plain} \nocite{*}
\bibliographystyle{plain}
\bibliography{biblio}

\begin{thebibliography}{10}

\bibitem{Beard20031012}
B.B. Beard.
\newblock Extending monte carlo samples.
\newblock {\em Nuclear Physics B - Proceedings Supplements}, 119:1012 -- 1014,
  2003.
\newblock Proceedings of the XXth International Symposium on Lattice Field
  Theory.

\bibitem{farhi-2009}
Edward Farhi, Jeffrey Goldstone, David Gosset, Sam Gutmann, Harvey~B. Meyer,
  and Peter Shor.
\newblock Quantum adiabatic algorithms, small gaps, and different paths, 2009.
\newblock arXiv:0909.4766.

\bibitem{farhi-2001}
Edward Farhi, Jeffrey Goldstone, Sam Gutmann, Joshua Lapan, Andrew Lundgren,
  and Daniel Preda.
\newblock A quantum adiabatic evolution algorithm applied to random instances
  of an \np-complete problem.
\newblock {\em Science}, 292:472--475, 2001.
\newblock arXiv:quant-ph/0104129.

\bibitem{krzakala-2008-78}
Florent Krzakala, Alberto Rosso, Guilhem Semerjian, and Francesco Zamponi.
\newblock On the path integral representation for quantum spin models and its
  application to the quantum cavity method and to monte carlo simulations.
\newblock {\em Physical Review B}, 78:134428, 2008.

\bibitem{Newman1999}
M.~Newman and G.~Barkema.
\newblock {\em Monte Carlo Methods in Statistical Physics}.
\newblock Clarendon Press, 1999.

\bibitem{prokofev-1996-64}
N.~V. Prokof'ev, B.~V. Svistunov, and I.~S. Tupitsyn.
\newblock Exact quantum monte carlo process for the statistics of discrete
  systems.
\newblock {\em ZH.EKS.TEOR.FIZ.}, 64:853, 1996.

\bibitem{0305-4470-25-13-017}
A~W Sandvik.
\newblock A generalization of handscomb's quantum monte carlo
  scheme-application to the 1d hubbard model.
\newblock {\em Journal of Physics A: Mathematical and General},
  25(13):3667--3682, 1992.

\bibitem{sandvik-1997-56}
A.~W. Sandvik, R.~R.~P. Singh, and D.~K. Campbell.
\newblock Quantum monte carlo in the interaction representation --- application
  to a spin-peierls model.
\newblock {\em Physical Review B}, 56:14510, 1997.

\bibitem{PTP.56.1454}
Masuo Suzuki.
\newblock Relationship between $d$-dimensional quantal spin systems and
  $(d+1)$-dimensional ising systems.
\newblock {\em Progress of Theoretical Physics}, 56(5):1454--1469, 1976.

\bibitem{wolff-2004-383}
Ulli Wolff.
\newblock Monte carlo errors with less errors.
\newblock {\em Comput.Phys.Commun.156:143-153,2004; Erratum-ibid.176:383,2007}.

\end{thebibliography}

\appendix

\section{Derivation of Estimators For the New Monte Carlo Method\label{sec:Derivation-of-Estimators}}

\subsection{Properties of the operator $A(E)$\label{sub:A(E)_properties}}

Our analysis of the new Monte Carlo method is based on properties
of the operator \[
A(E)=\left(\frac{-\lambda(E)}{H_{0}-E}V\right)\]
as was seen from the Introduction.

We take $E<0,$ and $\lambda(E)$ is defined to be the positive value
of $\lambda$ such that the ground state of $H(\lambda)$ has energy
$E$. We will show the following properties of this operator: 
\begin{enumerate}
\item All eigenvalues of $A(E)$ are real and $\leq1$ in absolute value.
\label{enu:property1-1} 
\item $|\psi_{g}(\lambda(E))\rangle$ is an eigenvector of $A(E)$ with
eigenvalue $+1.$ It may also be the case that $|\psi_{g}(-\lambda(E))\rangle$
is an eigenvector of $A(E)$ with eigenvalue $-1.$ There are no other
eigenvectors of $A(E)$ with eigenvalues $\pm1$. \label{enu:property2-1} 
\end{enumerate}
The second property says that the subspace of states spanned by eigenvectors
of $A(E)$ with $\pm1$ eigenvalues is either 1 or 2 dimensional.
We will see that the case where it is two dimensional occurs only
when $H(-\lambda(E))$ has ground state energy $E$.

To show that eigenvalues of $A(E)$ are real whenever $E<0$, note
that the spectrum of $A(E)$ is the same as the spectrum of the operator\[
\sqrt{H_{0}-E}A(E)\frac{1}{\sqrt{H_{0}-E}}\]
 which is Hermitian.

Now suppose (to reach a contradiction) that $|r\rangle$ is an eigenvector
of $A(E)$ with eigenvalue $R>1.$ Then \[
\bigg[H_{0}+\lambda(E)\frac{1}{R}V\bigg]|r\rangle=E|r\rangle\]
 which says that there exists an eigenvector of $H(\frac{\lambda(E)}{R})$
with eigenvalue $E$. In section \ref{sec:The-Hamiltonian} we proved
that the ground state energy of $H(\lambda)$ is strictly decreasing
for positive $\lambda$, which means that no eigenvector of $H(\frac{\lambda(E)}{R})$
can have energy smaller than or equal to $E$. This is a contradiction
and so all positive eigenvalues of $A(E)$ are $\leq1.$

Now, since we have proven that all eigenvalues of $A(E)$ are $\leq1$,
if it is the case that some negative eigenvalue of $A(E)$ is $<-1$
then the eigenvalue $W$ of $A(E)$ which is largest in magnitude
has negative sign. If this were true, then the limit

\[
\lim_{k\rightarrow\infty}\frac{1}{|W|^{2k+1}}Tr\bigg[(A(E))^{2k+1}\bigg]\]
 would be equal to a negative constant. But this cannot be the case
since all matrix elements of $A(E)$ are positive or zero. So we have
shown property 1.

We now proceed to show property 2 which was stated earlier in the
appendix, and in the process we prove the inequality \ref{eq:lambdaineq}
from section \ref{sec:The-Hamiltonian}. The real nonzero eigenvalues
of $A(E)$ can be related to eigenvalues of $H(\lambda)$ for some
value of $\lambda$. In particular, suppose that $\omega$ is a real
nonzero eigenvalue of $A(E)$ with eigenvector $|\omega\rangle$.
Then\begin{eqnarray*}
\left(H_{0}-E\right)\omega|\omega\rangle & = & -\lambda(E)V|\omega\rangle\end{eqnarray*}
 so

\[
H\left(\frac{\lambda(E)}{\omega}\right)|\omega\rangle=E|\omega\rangle\,.\]
Let us write the eigenvalues of $A(E)$ as \[
1=a_{1}(E)>a_{2}(E)\geq...\geq a_{2^{n}}(E)\geq-1\,.\]
 (which follows by property 1.) Then the values of $\lambda$ at which
$H(\lambda)$ has an eigenvalue with energy $E$ are\[
\frac{\lambda(E)}{a_{j}(E)}\]
 for $j\in\{1,...,2^{n}\}$. Write $\lambda_{-}(E)$ for the negative
value of lambda such that $H(\lambda_{-}(E))$ has ground state energy
$E$. The values of $\lambda$ which are smallest in magnitude (and
hence closest to the axis $\lambda=0$) are $\frac{\lambda(E)}{a_{1}(E)}=\lambda(E)>0$
and $\frac{\lambda(E)}{a_{2^{n}}(E)}=\lambda_{-}(E)<0$ and these
correspond to the ground state at energy $E$, with $|\lambda_{-}(E)|\geq\lambda(E)$.
All other values are greater than $\lambda(E)$ in magnitude. This
proves inequality \ref{eq:lambdaineq}, and also shows property 2
described above.

\subsection{Derivation of the Estimators $\beta_{est}$ and $\lambda_{est}^{2}$}

Having derived properties 1. and 2. of the operator $A(E)$ in the
previous section, we now proceed to use these properties to prove
equations \ref{eq:betaoverm} and \ref{eq:lim_lambda}.

Our treatment below applies to both the generic case where inequality
\ref{eq:lambdaineq} is strict as well as the nongeneric case where
equality holds as long as in the latter case $m$ is always even.
In either case we have \begin{equation}
Tr\left[(A(E))^{m}\right]=\left(\lambda(E)\right)^{m}F(E,m)\,.\label{eq:traceA}\end{equation}
 (Recall the definition of $F(E,m)$ from equation \ref{eq:def_F}.)
We can write this as\[
\left(\lambda(E)\right)^{m}F(E,m)=\sum_{i=1}^{2^{n}}\left(a_{i}(E)\right)^{m}\,.\]
 Taking the log and differentiating both sides gives

\begin{equation}
\frac{d}{dE}\left[\log\left(\lambda(E)^{m}\right)+\log\left(F(E,m)\right)\right]=\frac{1}{\sum_{i=1}^{2^{n}}\left(a_{i}(E)\right)^{m}}\sum_{j=1}^{2^{n}}m\left(a_{j}(E)\right)^{m-1}\frac{da_{j}}{dE}\,.\label{eq:difflog}\end{equation}
 Now take the limit as $m\rightarrow\infty$. Note that in the nongeneric
case the fact that $m$ is taken to be even ensures that the denominator
of equation \ref{eq:difflog} does not vanish in the limit of large
(even) $m$. The limit of the RHS is zero for any fixed value of $E$.
This is because in the large $m$ limit the only terms which contribute
to the sum in the numerator are those corresponding to values of $j$
for which $|a_{j}(E)|=1$. For these values of $a_{j}(E)$ (note $j$
is either $1$ or $2^{n}$ for these values) it is always the case
that $\frac{da_{j}}{dE}=0$ so these terms contribute zero to the
sum. 

So we have shown that \begin{equation}
\lim_{m\rightarrow\infty}\bigg(m\frac{1}{\lambda(E)}\frac{d\lambda}{dE}+\frac{1}{F(E,m)}\frac{dF(E,m)}{dE}\bigg)=0\,.\label{eq:firstderivnewmethod}\end{equation}
 Hence \begin{eqnarray*}
\lim_{m\rightarrow\infty}\frac{1}{m}\frac{1}{F(E,m)}\frac{dF(E,m)}{dE} & = & -\frac{1}{\lambda(E)}\frac{d\lambda}{dE}.\end{eqnarray*}
 The left hand side of this equation can be rewritten as an ensemble
average, which results in \begin{equation}
\lim_{m\rightarrow\infty}\frac{1}{m}\frac{1}{F(E,m)}\frac{dF(E,m)}{dE}=\lim_{m\rightarrow\infty}\langle\frac{1}{m}\sum_{i=1}^{m}\frac{1}{E_{i}-E}\rangle_{f}=-\frac{1}{\lambda(E)}\frac{d\lambda}{dE}\,.\label{eq:estimator1}\end{equation}
 So for fixed $m$ sufficiently large, one can estimate the quantity
$-\frac{1}{\lambda(E)}\frac{d\lambda}{dE}$ 
using the ensemble average $\frac{\bar{\beta}(E,m)}{m}$. This proves
equation \ref{eq:betaoverm}.

We now derive an estimator for the quantity $\lambda(E)$ itself as
an ensemble average. 
For this purpose we make use of the fact that \[
\lim_{m\rightarrow\infty}\frac{F(E,m-2)}{F(E,m)}=\lambda(E)^{2}\,.\]
 Expanding the numerator and denominator as sums over paths, we obtain
\begin{equation}
\frac{F(E,m-2)}{F(E,m)}=\frac{\sum_{\{z_{1},...,z_{m-2}\}}\langle z_{1}|-V|z_{m-2}\rangle...\langle z_{2}|-V|z_{1}\rangle\prod_{i=1}^{m-2}\frac{1}{E_{i}-E}}{\sum_{\{z_{1},...,z_{m}\}}\langle z_{1}|-V|z_{m}\rangle...\langle z_{2}|-V|z_{1}\rangle\prod_{i=1}^{m}\frac{1}{E_{i}-E}}\,.\label{eq:Fratio}\end{equation}
 Now rewrite the numerator as \begin{eqnarray*}
F(E,m-2) & = & \sum_{\{z_{1},...,z_{m}\}}\bigg(\langle z_{1}|-V|z_{m}\rangle\langle z_{m}|-V|z_{m-1}\rangle\langle z_{m-1}|-V|z_{m-2}\rangle...\langle z_{2}|-V|z_{1}\rangle\\
 &  & \prod_{i=1}^{m}\frac{1}{E_{i}-E}\left[\delta_{z_{1}z_{m-1}}(E_{m}-E)(E_{1}-E)\frac{1}{\langle z_{1}|V^{2}|z_{1}\rangle}\right]\bigg)\end{eqnarray*}
 where $\delta_{z_{1}z_{m-1}}$ is the Kronecker delta. Since our
distribution $f(z_{1},...,z_{m})$ is invariant under cyclic permutations
of the bit strings $\{z_{i}\}$, we can write\begin{eqnarray*}
F(E,m-2) & = & \sum_{\{z_{1},...,z_{m}\}}\bigg(\langle z_{1}|-V|z_{m}\rangle\langle z_{m}|-V|z_{m-1}\rangle\langle z_{m-1}|-V|z_{m-2}\rangle...\langle z_{2}|-V|z_{1}\rangle\\
 &  & \prod_{j=1}^{m}\frac{1}{E_{j}-E}\left[\frac{1}{m}\sum_{i=1}^{m}\delta_{z_{i+2}z_{i}}(E_{i+1}-E)(E_{i}-E)\frac{1}{\langle z_{i}|V^{2}|z_{i}\rangle}\right]\bigg)\,\end{eqnarray*}
where $z_{m+1}=z_{1}$ and $z_{m+2}=z_{2}$. Inserting this formula
into equation \ref{eq:Fratio} gives the final expression for $\lambda(E)^{2}$
as an ensemble average \begin{equation}
\lim_{m\rightarrow\infty}\bigg<\frac{1}{m}\sum_{i=1}^{m}\delta_{z_{i+2}z_{i}}(E_{i+1}-E)(E_{i}-E)\frac{1}{\langle z_{i}|V^{2}|z_{i}\rangle}\bigg>_{f}=\lambda(E)^{2}\,.\label{eq:estimator2}\end{equation}
 This proves equation \ref{eq:lim_lambda}.

\section{Estimators in Continuous Imaginary Time Quantum Monte Carlo\label{sec:Estimators-in-Continuous}}

In this section we derive the known estimators for $\langle H_{0}\rangle$
and $\langle\lambda V\rangle$ stated in equations \ref{eq:standardestimators}
and \ref{eq:standardV}.

\subsection*{Estimator for $\langle H_{0}\rangle$ }

To derive the estimator for $\langle H_{0}\rangle$ we write (using
the Dyson series to expand $e^{-\beta H})$

\[
\frac{Tr[H_{0}e^{-\beta H}]}{Tr[e^{-\beta H}]}=\frac{1}{Z(\beta)}Tr\bigg[H_{0}\sum_{m=0}^{\infty}(-\lambda)^{m}e^{-\beta H_{0}}\int_{0}^{\beta}dt_{m}\int_{0}^{t_{m}}dt_{m-1}...\int_{0}^{t2}dt_{1}V_{I}(t_{m})V_{I}(t_{m-1})...V_{I}(t_{1})\bigg]\]
 where $V_{I}(t)=e^{tH_{0}}Ve^{-tH_{0}}.$ (The $m=0$ term in the
above sum is $\frac{1}{Z(\beta)}Tr\big[e^{-\beta H_{0}}\big]$.) Inserting
complete sets of states in the basis $\{|z\rangle\}$ which diagonalizes
$H_{0}$ we obtain \begin{eqnarray*}
\frac{Tr[H_{0}e^{-\beta H}]}{Tr[e^{-\beta H}]} & = & \frac{1}{Z(\beta)}\sum_{m=0}^{\infty}\bigg[(-\lambda)^{m}\sum_{\{z_{1},...,z_{m}\}}\langle z_{1}|H_{0}|z_{1}\rangle\langle z_{1}|V|z_{m}\rangle\langle z_{m}|V|z_{m-1}\rangle...\langle z_{2}|V|z_{1}\rangle\\
 &  & \int_{0}^{\beta}dt_{m}\int_{0}^{t_{m}}dt_{m-1}...\int_{0}^{t_{2}}dt_{1}e^{-(E_{1}t_{1}+E_{2}(t_{2}-t_{1})+...+E_{1}(\beta-t_{m}))}\bigg]\\
 & = & \langle H_{0}(z(t=0))\rangle_{\rho}\,.\end{eqnarray*}
 where the expectation value is with respect to the measure $\rho$
defined in section \ref{sec:Standard}, and $H_{0}(z(t=0))=\langle z(0)|H_{0}|z(0)\rangle.$
Noting that the measure $\rho$ is invariant under a translation of
the path by a time $x\in[0,\beta]$ (this corresponds to the transformation
$t_{i}$ goes to $(t_{i}+x\text{) mod \ensuremath{\beta}}$ for $i\in1,...,m$
followed by a reordering of the labels $i$ to maintain time ordering),
we have that\[
\langle H_{0}(z(t=0))\rangle_{\rho}=\langle H_{0}(z(t=x))\rangle_{\rho}\text{ , \ensuremath{}for all x\ensuremath{\in}[0,\ensuremath{\beta}]. }\]
 We obtain the stated estimator for $\langle H_{0}\rangle$ by averaging
over all $x\in[0,\beta]$ \[
\langle H_{0}\rangle\equiv\frac{Tr[H_{0}e^{-\beta H}]}{Tr[e^{-\beta H}]}=\langle\frac{1}{\beta}\int_{0}^{\beta}H_{0}(z(x))dx\rangle_{\rho.}\,.\]

\subsection*{Estimator for $\langle\lambda V\rangle$}

As in the previous section, we begin by expanding the operator $e^{-\beta H}$
\begin{equation}
\frac{Tr[\lambda Ve^{-\beta H}]}{Tr[e^{-\beta H}]}=\frac{1}{Z(\beta)}Tr\bigg[\lambda V\sum_{m=0}^{\infty}(-\lambda)^{m}e^{-\beta H_{0}}\int_{0}^{\beta}dt_{m}\int_{0}^{t_{m}}dt_{m-1}...\int_{0}^{t2}dt_{1}V_{I}(t_{m})V_{I}(t_{m-1})...V_{I}(t_{1})\bigg]\,.\label{eq:Vseries}\end{equation}
 For $m=0,1,2...$ we have\begin{eqnarray*}
\int_{0}^{\beta}dt_{m+1}\int_{0}^{t_{m+1}}dt_{m}...\int_{0}^{t2}dt_{1}V_{I}(t_{m+1})V_{I}(t_{m})...V_{I}(t_{1})\delta(t_{1})\\
=\int_{0}^{\beta}dt_{m+1}\int_{0}^{t_{m+1}}dt_{m}...\int_{0}^{t3}dt_{2}V_{I}(t_{m+1})V_{I}(t_{m})...V_{I}(t_{2})V\,\\
=\int_{0}^{\beta}dt_{m}\int_{0}^{t_{m}}dt_{m-1}...\int_{0}^{t2}dt_{1}V_{I}(t_{m})V_{I}(t_{m-1})...V_{I}(t_{1})V\,.\end{eqnarray*}
 Plugging this expression into \ref{eq:Vseries} we obtain \begin{eqnarray*}
\frac{Tr[\lambda Ve^{-\beta H}]}{Tr[e^{-\beta H}]} & = & \frac{1}{Z(\beta)}Tr\bigg[(-1)\sum_{m=0}^{\infty}(-\lambda)^{m+1}e^{-\beta H_{0}}\\
 &  & \int_{0}^{\beta}dt_{m+1}\int_{0}^{t_{m+1}}dt_{m}...\int_{0}^{t2}dt_{1}V_{I}(t_{m+1})V_{I}(t_{m})...V_{I}(t_{1})\delta(t_{1})\bigg]\\
 & = & \frac{1}{Z(\beta)}Tr\bigg[(-1)\sum_{m=1}^{\infty}(-\lambda)^{m}e^{-\beta H_{0}}\\
 &  & \int_{0}^{\beta}dt_{m}\int_{0}^{t_{m}}dt_{m-1}...\int_{0}^{t2}dt_{1}V_{I}(t_{m})V_{I}(t_{m-1})...V_{I}(t_{1})\delta(t_{1})\bigg]\\
 & = & -\langle(1-\delta_{m,0})\delta(t_{1})\rangle_{\rho}\\
 & = & -\langle(1-\delta_{m,0})\sum_{l=1}^{m}\delta(t_{l})\rangle_{\rho}\quad\text{(since only }t_{1}\text{ can ever be \mbox{0).}}\end{eqnarray*}
 In the last two lines of the above $m$ appears inside an expectation
value $\langle...\rangle_{\rho}$. In this context $m$ is considered
to be a function of the path. Now we can use the fact that the measure
$\rho$ over paths is invariant under translations of the path in
imaginary time to write\begin{eqnarray*}
\langle\lambda V\rangle\equiv\frac{Tr[\lambda Ve^{-\beta H}]}{Tr[e^{-\beta H}]} & = & -\langle(1-\delta_{m,0})\frac{1}{\beta}\int_{0}^{\beta}\sum_{l=1}^{m}\delta(t_{l}-t)dt\rangle_{\rho}\\
 & = & -\langle\frac{m}{\beta}\rangle_{\rho}\,.\end{eqnarray*}
 So $\langle\lambda V\rangle$ is $-\frac{1}{\beta}$ times the average
number of jumps in a path of length $\beta$.

\section{Convergence of The Markov Chain\label{sec:Ergodicity-and-Detailed}}

We show in this section that the Markov Chain defined in section \ref{sec:An-Example:-Transverse}
can be used to estimate any quantity which is invariant under cyclic
permutations of the path. In order to streamline the proof, it will
be useful to define a different Markov Chain over paths which has
the update rule (for some fixed $0<p<1$)
\begin{enumerate}
\item With probability $p$ do 1 update of the Markov Chain defined in section
\ref{sec:An-Example:-Transverse}. 
\item With probability $1-p$ apply a random cyclic permutation to the path
by letting $\{z_{1},z_{2},...,z_{m}\}\rightarrow\{z_{j},z_{j+1},...,z_{m},z_{1},...z_{j-1}\}$
for uniformly random $j\in\{1,...,m\}$. 
\end{enumerate}
In the next two sections we show that the above Markov Chain has limiting
distribution $f$ (defined in equation \ref{eq:little_f}) for any
choice of the parameter $0<p<1$. This Markov Chain with fixed $0<p<1$
induces a random walk on equivalence classes of paths where an equivalence
class is the set of all paths related to a given path by cyclic permutation.
In this equivalence class random walk, step 2 does nothing. So the
limiting distribution over equivalence classes is the same whether
or not step 2 is performed. If one only estimates quantities which
are invariant under cyclic permutations (note that the estimators
we have discussed have this property) then one can use the algorithm
with $p=1$ (so step 2 is never performed).

To show that the above Markov Chain converges to the limiting distribution
$f$ over paths, it is sufficient to verify that the update rule constructed
above satisfies the following two conditions \cite{Newman1999}: 
\begin{itemize}
\item \emph{Ergodicity}: Given any two paths $A$ and $B$, it is possible
to reach path $B$ by starting in path $A$ and applying the Markov
chain update rule a finite number of times. 
\item \emph{Detailed Balance:} For any two paths $A$ and $B$, \begin{equation}
f(A)P(A\rightarrow B)=f(B)P(B\rightarrow A)\label{eq:detailedbalance}\end{equation}
 where $P(X\rightarrow Y$) is the probability of transitioning to
the path $Y$ given that you start in path $X$ and apply one step
of the Markov chain. 
\end{itemize}
We now show that this Markov Chain satisfies these conditions.

\subsection*{Ergodicity}

In order to show ergodicity of the Markov Chain defined above, we
first note that a path can be specified either by a list of bit strings
$\{z_{1},...,z_{m}\}$, or by one bit string $z_{start}$ followed
by a list of bits in which flips occur $\{b_{1},...,b_{m}\}$, with
each $b_{r}\in\{1,...,n\}$.

We now show that by applying the Monte Carlo update rules illustrated
in figures \ref{fig:interchange_update} and \ref{fig:replace_update},
it is possible to transform an arbitrary path $A\longleftrightarrow\{z_{start},\{b_{1},...,b_{m}\}\}$
into another arbitrary path $B\longleftrightarrow\{y_{start,}\{c_{1},...,c_{m}\}\}$
where $y_{start}$ and $z_{start}$ differ by an even number of bit
flips. This is sufficient to show ergodicity because any path $B$
can be cyclically permuted into a path which starts in a state $\tilde{y}_{start}$
that differs from $z_{start}$ by an even number of flips (and our
Markov chain includes moves which cyclically permute the path). We
assume here that $m\geq4$, since we are interested in the limit of
large m anyways. In order to transform path $A$ into path $B$, we
give the following prescription: 
\begin{enumerate}
\item First transform $z_{start}$ into $y_{start}$. To do this, note that
one can move any two flips $b_{i}$ and $b_{j}$ so that they are
just before and just after $z_{start}$ (i.e $\tilde{b}_{1}=b_{j}$
and $\tilde{b}_{m}=b_{i})$, by applying the flip interchange rule
illustrated in figure \ref{fig:interchange_update}. So long as you
do not interchange the first and last flip in the list, the bit string
$z_{start}$ will remain unchanged. Then one can flip both of these
bits in the bit string $z_{start}$ by interchanging the two flips
$\tilde{b}_{1}$ and $\tilde{b}_{m}.$ This describes how to flip
any two bits in $z_{start}$, assuming that flips in these bits occur
somewhere in the path. Now suppose that you wish to flip 2 bits in
$z_{start}$ but one or both of the bits does not occur in the current
list of flips in the path. In that case you must first take some pair
of flips which occur in some other bit $q$, and then move them until
they are adjacent using the flip interchange rule (without ever moving
them past $z_{start}).$ Once they are adjacent, you can replace them
with a pair of flips in another bit using the flip replacement rule
illustrated in figure \ref{fig:replace_update} . Assuming $m\geq4,$
there will always be two pairs of flips in the path which can be replaced
by flips in the two bits that you desire to change in $z_{start}.$ 
\item After $z_{start}$ has been transformed into $y_{start},$ one must
then make the list of flips equal to $\{c_{1},...,c_{m}\}.$ This
can be done by interchanging flips and replacing pairs of flips as
described above. Since this can always be achieved without interchanging
the first and last flip in the path, the bit string $y_{start}$ will
remain unchanged by this procedure. 
\end{enumerate}

\subsection*{Detailed Balance}

Here we demonstrate that the Markov Chain defined above satisfies
the detailed balance condition from equation \ref{eq:detailedbalance}.
To show this, fix two paths $A$ and $B$ and consider the probability
of transitioning between them in one step of the Monte Carlo update
rule. This probability is zero except in the following cases 
\begin{enumerate}
\item $A=B$. In this case detailed balance is trivially satisfied.
\item A cyclic permutation of the bit strings (which is not the identity)
maps the path $A$ into the path $B$. In this case $f(A)=f(B)$ and
the probability of transitioning from $A$ to $B$ in one move of
the Markov Chain is also equal to the probability of the reverse transition
from $B$ to $A$ (this is because for every cyclic permutation which
maps $A$ to $B$ the inverse permutation is also cyclic and maps
$B$ to $A$). So detailed balance is satisfied.
\item $A$ and $B$ are the same path except for at one location. In other
words, $A$ can be described by the sequence $\{z_{1}^{A},...,z_{m}^{A}\}$
and $B$ can be described by the sequence $\{z_{1}^{B},...,z_{m}^{B}\}$
where the corresponding bit strings are all the same except for one
pair $z_{i}^{A}$ and $z_{i}^{B}$. Write $q_{1}$ for the bit in
which $z_{i}^{A}$ and $z_{i-1}^{A}$ differ, and $q_{2}$ for the
bit in which $z_{i}^{A}$ and $z_{i+1}^{A}$ differ. We have to consider
two cases depending on whether or not $q_{1}=q_{2}:$\end{enumerate}
\begin{itemize}
\item Case 1: $q_{1}\neq q_{2}$. Detailed balance follows in this case
since the transition probabilities follow the Metropolis Monte Carlo
rule\begin{eqnarray*}
\frac{P(A\rightarrow B)}{P(B\rightarrow A)} & = & \min\left\{ 1,\frac{E_{i}^{A}-E}{E_{i}^{B}-E}\right\} \frac{1}{\min\left\{ 1,\frac{E_{i}^{B}-E}{E_{i}^{A}-E}\right\} }\\
\frac{f(A)}{f(B)} & = & \frac{E_{i}^{B}-E}{E_{i}^{A}-E}=\frac{P(B\rightarrow A)}{P(A\rightarrow B)}\,.\end{eqnarray*}

\item Case 2: $q_{1}=q_{2}$. In this case, from equation \ref{eq:same_q}
we have \begin{eqnarray*}
\frac{P(A\rightarrow B)}{P(B\rightarrow A)} & = & \frac{E_{i}^{A}-E}{E_{i}^{B}-E}\end{eqnarray*}
 and \[
\frac{f(A)}{f(B)}=\frac{E_{i}^{B}-E}{E_{i}^{A}-E}=\frac{P(B\rightarrow A)}{P(A\rightarrow B)}\,.\]

\end{itemize}

\end{document}